# Viral Dark Matter: Illuminating Protein Function, Ecology, and Biotechnological Promises

James C. Kosmopoulos[1,2] & Karthik Anantharaman[1,3,4*]

[1] Department of Bacteriology, University of Wisconsin-Madison, Madison, WI, USA
[2] Microbiology Doctoral Training Program, University of Wisconsin-Madison, Madison, WI, USA
[3] Department of Integrative Biology, University of Wisconsin-Madison, Madison, WI, USA
[4] Department of Data Science and AI, Wadhwani School of Data Science and AI, Indian Institute of Technology Madras, Chennai, Tamil Nadu, India
* Correspondence should be addressed to KA

## Abstract

Viruses are the most abundant biological entities on Earth and play central roles in shaping microbiomes and influencing ecosystem functions. Yet, most viral genes remain uncharacterized, comprising what is commonly referred to as "viral dark matter." Metagenomic studies across diverse environments consistently show that 40-90% of viral genes lack known homologs or annotated functions. This persistent knowledge gap limits our ability to interpret viral sequence data, understand virus-host interactions, and assess the ecological or applied significance of viral genes. Among the most intriguing components of viral dark matter are auxiliary viral genes (AVGs), including auxiliary metabolic genes (AMGs), regulatory genes (AReGs), and host physiology-modifying genes (APGs), which may alter host function during infection and contribute to microbial metabolism, stress tolerance, or resistance. In this review, we explore recent advances in the discovery and functional characterization of viral dark matter. We highlight representative examples of novel viral proteins across diverse ecosystems including human microbiomes, soil, oceans, and extreme environments, and discuss what is known, and still unknown, about their roles. We then examine the bioinformatic and experimental challenges that hinder functional characterization, and present emerging strategies to overcome these barriers. Finally, we highlight both the fundamental and applied benefits that multidisciplinary efforts to characterize viral proteins can bring. By integrating computational predictions with experimental validation, and fostering collaboration across disciplines, we emphasize that illuminating viral dark matter is both feasible and essential for advancing microbial ecology and unlocking new tools for biotechnology.

## 1. Introduction

Viruses are the most numerous biological entities on Earth, infecting organisms across all domains of life and shaping the structure and function of virtually every ecosystem[1–3]. Among them, bacteriophages (phages; viruses that infect bacteria) are most abundant in nature. Phages manipulate their hosts over the course of infection with profound consequences on microbiomes[4]. This manipulation is often achieved by auxiliary viral genes (AVGs), which are genes not essential for viral replication but which augment host metabolic (AMGs), physiological (APGs), and regulatory processes (AReGs)[5]. When expressed, AVGs "reprogram" key host functions to ultimately benefit phage reproduction[6–8]. At the cellular scale, AVGs may boost energy production to support viral genome replication[8–10], activate toxin-antitoxin systems that suppress competing phages[11,12], or inhibit host sporulation and dormancy to maintain favorable conditions for infection[13]. When such interactions scale across microbial communities, they can drive global shifts in ecosystem function and microbial evolution[14–18]. Yet, beyond a limited number of well-characterized examples, we do not know the diversity of ways in which phage manipulate their hosts due to our lack of ability to annotate viral proteins.

Owing to viral genomic diversity and rapid evolution, our ability to assign functions to viral proteins by sequence similarity is severely limited. Environmental surveys consistently show that a large fraction of viral genes lack any functional annotation. In environmental studies, 40–90% of viral DNA sequences cannot be assigned to known functions or even align to previously described viral sequences[19–22], a phenomenon often termed "viral dark matter." Even in curated databases of reference viral genomes, roughly 40–45% of proteins[23,24] and 75–85% of protein families are annotated as hypothetical or unknown (Figure 1). The gap between viral gene discovery and functional characterization is widening: modern metagenomic studies are uncovering millions of new viral genes and genomes, yet most of these encode proteins of unknown function[25]. For example, the IMG/VR v4 database now contains >15 million viral sequences[25], and a recent human gut virus catalog added >450,000 new viral protein clusters (92% previously undocumented)[26]. This accumulation of viral proteins with undefined roles represents a major bottleneck in understanding virus-host interactions. Crucially, most viral gene functions inferred from sequence remain putative until experimentally validated, leaving many predicted roles unconfirmed in the lab.

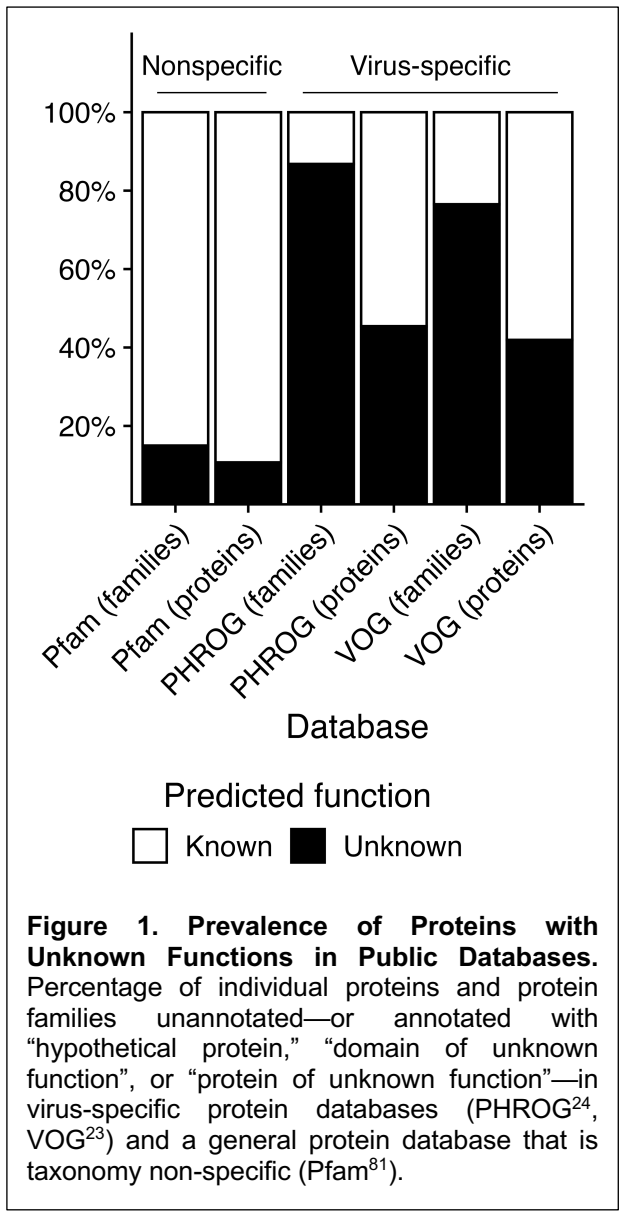

**Figure 1. Prevalence of Proteins with Unknown Functions in Public Databases.** Percentage of individual proteins and protein families unannotated—or annotated with "hypothetical protein," "domain of unknown function", or "protein of unknown function"—in virus-specific protein databases (PHROG[24], VOG[23]) and a general protein database that is taxonomy non-specific (Pfam[81]).

Why is viral "dark matter" so pervasive? A key reason is that the majority of microbes (and thus their viruses which are obligate parasites) cannot be readily cultured in the lab. For cellular microbes, metagenomic sequencing has revealed that cultivated isolates represent under 10% of total microbial biodiversity, with the rest coming from uncultivated lineages[27]. Viruses face a similar gap – the vast majority of known viral diversity has been uncovered by cultivation-independent methods only[25]. Additionally, viral genomes and proteins often evolve rapidly and share little sequence similarity to known references, hindering homology-based annotations[22]. Many viral genes are small or novel "open reading frames" (ORFs) with weakly conserved domains[28,29], and some functions may be context-dependent (e.g. acting only under specific host or environmental conditions)[30,31], making them hard to predict computationally. As a result, standard genome annotation pipelines assign generic labels like "hypothetical protein" to roughly half or more of viral genes in a novel genome sequence[23,24]. Overcoming this functional unknown is critical: without understanding what these viral proteins do, we miss key insights into viral ecology, their roles in microbial metabolism, and potential applications in phage therapy and biotechnology.

Characterizing the functions of these unknown viral proteins holds considerable promise for both ecological understanding and practical applications. From an ecological standpoint, viruses are pivotal regulators of microbial populations and nutrient transformations[2,4]. Uncharacterized viral genes could be mediating metabolic processes like carbon, nitrogen and phosphorus transformations, or host–microbe interactions in ways we have yet to recognize. Every new function illuminated within viral dark matter can reveal novel mechanisms by which viruses impact host metabolism or shape ecosystem dynamics. From an applied perspective, viral genomes represent a vast reservoir of novel enzymes and bioactive molecules. Indeed, phage-derived proteins have already provided invaluable tools in biotechnology (for instance, various DNA polymerases and ligases used in molecular biology originate from phage genes)[32] and potent antibacterial agents (e.g. phage lytic enzymes in phage therapy)[33,34]. It stands to reason that many of the currently unknown viral proteins could similarly be harnessed for biotechnological innovation or as therapeutics. In short, shining light on viral dark matter will deepen our understanding of how viruses are microbial ecosystem engineers, and may also uncover new enzymes for bioengineering and medicine.

In this review, we examine the emerging efforts to illuminate viral dark matter – the vast repertoire of viral proteins with unknown functions. First, we describe how modern 'omics techniques have vastly expanded the known virosphere and provided initial clues to viral gene functions even without culturing. Next, we highlight representative examples of newly discovered viral metabolic genes and other enigmatic viral proteins across a range of environments – from the human gut, to soil, oceans, and extreme habitats, to illustrate their ecological significance. We then discuss the major obstacles to functional characterization of viral genes, including bioinformatic limitations and experimental challenges, to understand why bridging this knowledge gap is so difficult. Finally, we explore emerging strategies and technologies aimed at overcoming these hurdles – emphasizing cross-disciplinary approaches that connect computational predictions with laboratory validation. By synthesizing what is "known about the unknowns" for viruses and outlining a roadmap for their characterization, our aim is to accelerate the integration of viral dark matter into more biochemical studies to broaden our collective understanding of microbiomes.

## 2. 'Omics-Based Discovery of Viral Dark Matter

Early efforts to mine microbial genomes had identified thousands of viral genomes integrated in bacterial/archaeal genomes (prophages), providing the first viral representatives for dozens of new microbial phyla[19]. In one early study, Roux *et al.* (2015) recovered ~12,500 viral genomes from publicly available microbial genomes, including viruses infecting 13 bacterial phyla that were previously unsampled at the time[19]. Later efforts focused on discovering viral genomes not just from cultivated bacterial/archaeal genomes, but from entire microbial communities (metagenomics). By directly sequencing DNA from environmental or host-associated samples, metagenomics can uncover an enormous diversity of viruses (Table 1), including many that infect uncultivated microbes. Public databases of viral genomes have since exploded in size, for example, the IMG/VR v4 database published in 2023 now contains >15 million viral genomes or genome fragments derived largely from metagenomes[25]. This is a six-fold increase over the previous release from 2021[35], highlighting the explosion of new viral sequences discovered from metagenomics in a relatively short period of time. Likewise, metagenomic surveys of the human, soil, ocean, and other habitats are continually expanding the known virosphere. For instance, recent human gut virome studies alone have each discovered tens or hundreds of thousands of viral genomes, many with no close matches in existing virus databases[26,36–38]. Metagenomics has thus revolutionized the discovery of viruses and their genes, bypassing the need for culturing hosts or viruses.

**Table 1. 'Omics-Based Methods Used to Study Viral Proteins, their Benefits, and Drawbacks.**

| Method | Starting Molecule | Benefits | Drawbacks |
| --- | --- | --- | --- |
| **Metagenomics** | DNA (environmental or host-associated DNA) | - Captures **viral genomes** directly from samples without need for culturing hosts or viruses, enabling discovery of vast numbers of new viruses (including those infecting uncultivated microbes).<br><br>- Reveals the **genetic capacity** of viruses: uncovers viral genes that may hint at virus-host interactions. | - **Biased to DNA viruses:** Fails to detect RNA viruses, necessitating metatranscriptomics for RNA virus discovery<br><br>- **Provides limited insight into activity:** DNA presence doesn't indicate if a viral gene is expressed or functional in the environment. |
| **Metatranscriptomics** | RNA (total community RNA, | - Detects **actively expressed viral genes** as RNA transcripts, | - **RNA is less stable**: Viral RNA can be rare and easily degraded; samples often |

| | often after rRNA depletion) | highlighting which viral functions are in use *in situ*<br><br>- Enables discovery of **RNA viruses** that lack DNA stages and are invisible to DNA metagenomics. | require careful processing and enrichment.<br><br>- **Biased toward current infections:** Only detects viruses that are actively transcribing. Dormant viruses or DNA viruses with no ongoing transcription in the sample will be missed, potentially underestimating total viral contributions. |
|---|---|---|---|
| **Metaproteomics** | Proteins (all proteins extracted from a community sample) | - **Confirms protein-level expression of viral genes**: Verifies that hypothetical ORFs from viral genomes are translated and allows functional inferences.<br><br>- **Complements genomic data** by assigning tentative functions to unknown viral proteins based on detected peptides. | - **Low sensitivity for viruses:** viral proteins are often low-abundance amid a vast host protein background, so metaproteomics may preferentially detect only the most abundant viral proteins.<br><br>- **Technical complexity:** environmental proteomics is experimentally and computationally demanding, needing extensive sample processing, high-end mass spectrometry, and complex peptide-to-protein matching pipelines. |

While gene catalogs from metagenomics outline the scope of viral dark matter, complementary 'omics approaches can add an additional dimension by identifying which viral genes are active in nature, and by identifying uncultivated RNA viruses. Metatranscriptomics (the bulk sequencing of RNA from entire microbiomes) has emerged as a powerful tool for discovering RNA virus genomes (Table 1), providing insights into their encoded genes and potential ecological roles that remain elusive from DNA-based studies alone. Recent metatranscriptomic analyses have significantly expanded our understanding of RNA virus diversity, particularly in understudied soil environments. For instance, Starr *et al.* (2019) employed metatranscriptomics to reconstruct the RNA viral community across multiple soil habitats, uncovering thousands of novel RNA viruses primarily from the *Narnaviridae* and *Leviviridae* families[39]. Similarly, Hillary *et al*. (2022) identified thousands of RNA viral sequences across different grassland soils using RNA viromics (metatranscriptomics enriched for virus-like particles), demonstrating the prevalence of diverse RNA viruses infecting not only bacteria but also fungi, plants, vertebrates, and invertebrates[40]. Notably, a significant fraction of these viruses belonged to previously understudied groups. Similarly, Wu *et al.* (2022) and Pratama *et al.* (2025) utilized metatranscriptomics to explore RNA viruses in permafrost soils undergoing thaw due to climate change, with each study revealing thousands of novel RNA viruses encoding AVGs potentially involved in nutrient transformations and host metabolism[41,42]. Together, these studies emphasize RNA viruses as key yet largely uncharacterized contributors to viral dark matter, underscoring the need for continued functional investigations into RNA-virus encoded proteins.

Going a step further from DNA- and RNA-based inferences, direct analyses of translated viral proteins have been particularly illuminating. Metaproteomics (mass-spectrometry-based protein identification in environmental samples) has been used to detect microbially encoded proteins in complex communities (Table 1), confirming the production of many hypothetical proteins and generating new hypotheses[43,44]. A landmark study by Brum *et al.* (2016) applied metaproteomics alongside DNA sequencing to ocean water samples, identifying 1,875 virus-associated structural proteins from uncultivated marine viruses[20]. Remarkably, over half of these proteins had been previously unannotated but were assigned broad functional categories through their approach, such as "capsid protein" or "tail protein". The most abundant proteins in their datasets turned out to be components of viral capsids, suggesting that a particular conserved capsid fold may be among the most abundant biological structures known to science[20]. More metaproteomic studies of that included viruses in human[43], ruminant[45], and soil[46] microbiomes discovered more previously unknown virus-encoded proteins with metabolic functions that have the potential to impact their entire communities. Collectively, these studies demonstrated that proteomics could illuminate portions of the viral dark matter: by directly observing proteins, researchers proposed functions for hundreds of widespread and conserved viral genes that were previously unknown sequences, and were thus able to propose new frameworks for microbial and viral functions in microbiomes.

Together, culture-independent metagenomics, metatranscriptomics, and metaproteomics have dramatically improved our understanding of the functional diversity of viruses. They enable high-throughput discovery of novel genomes, genes, and proteins that guide further study. Still, although assigning potential functions for proteins encoded by uncultivated viruses have been a big leap, specific biochemical activities of these proteins remained unverified in many cases, highlighting the need for deeper functional analysis to move from functional predictions to proof.

## 3. Viral Dark Matter Across Diverse Ecosystems

### 3.1. Human Gut Microbiome

Viruses are integral to human microbiomes, with the gut virome being particularly rich in phages. Large-scale metagenomic projects have revealed tens of thousands of distinct viral species in the human gut, most of which were previously unknown[26,36–38]. A 2021 study of human gut viruses by Nayfach *et al*. clustered ~11.8 million viral genes into 459,375 protein clusters, finding that 45% of the genes had no matches to any known profiles and another 30% matched only profiles of unknown function[26]. In other words, ~75% of human gut viral genes currently lack a clear functional annotation. Interestingly, only 39% of the 459,375 protein clusters were singleton proteins, suggesting that the remaining 61% have homologs in the human gut. This means that although most viral genes in the human gut have no currently known function, they are still abundant and conserved, suggesting that they are functionally important to the human gut microbiome overall and make them promising candidates that warrant further study. Of the proteins that were able to be assigned putative functions, some of the largest protein clusters included phage structural proteins, DNA packaging, replication, and binding proteins, lysis proteins, and interestingly, reverse transcriptases, which suggested that diversity generating retroelements (DGRs) are highly prevalent in the human gut and contribute to virus-host coevolutionary dynamics. Some beta-lactamases were also found to be encoded by human gut viruses, suggesting that a portion of viral dark matter in the human gut may be involved in antibiotic resistance, although this is thought to be a rare phenomenon[47].

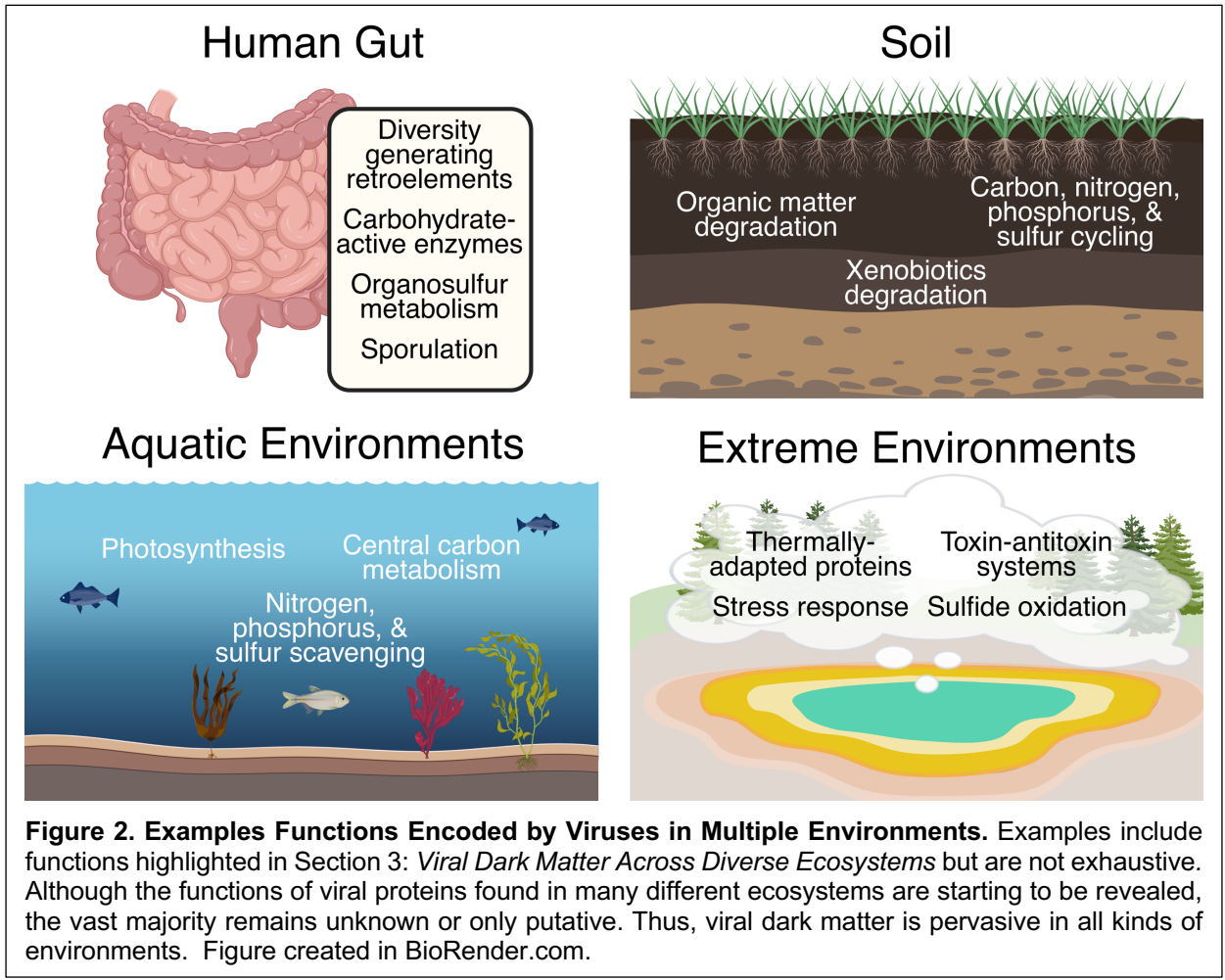

**Figure 2. Examples Functions Encoded by Viruses in Multiple Environments.** Examples include functions highlighted in Section 3: *Viral Dark Matter Across Diverse Ecosystems* but are not exhaustive. Although the functions of viral proteins found in many different ecosystems are starting to be revealed, the vast majority remains unknown or only putative. Thus, viral dark matter is pervasive in all kinds of environments. Figure created in BioRender.com.

Despite the large fraction of viral proteins in the human gut with unknown functions, some metabolic or ecological roles of gut phage genes are beginning to emerge from metagenomic data (Figure 2). Bacteriophages in the gut have been found to carry auxiliary genes that could influence bacterial host fitness or metabolism. For instance, gut phages sometimes encode auxiliary carbohydrate-active enzymes that might help their hosts degrade mucin in the gut lining or dietary polysaccharides[48]. In another study by Kieft *et al*. (2021), viruses encoding enzymes involved in assimilatory and organic sulfur metabolism were ubiquitous in many environments, especially the human gut[49]. They also experimentally demonstrated that increasing concentrations of sulfide ($H_2S$), a byproduct of the viral sulfate reduction enzymes, was met with an increase in viral population sizes, which can have far reaching implications on the overall gut microbiome and human health[49]. Beyond metabolism, human gut phages may influence host life cycles. A recent study noted that human gut phages harbor genes involved in bacterial sporulation[50], hinting that phages might induce or modulate spore formation in their hosts. However, in most of the cases highlighted here, the actual activity of the phage-encoded proteins remains to be demonstrated experimentally. The human gut virome thus contains a myriad of novel genes that potentially affect host physiology, gene regulation, and metabolism, but confirming these functions still requires targeted experiments.

## 3.2. Soil Environments

Soils harbor some of the most complex viral communities. Metagenomic surveys of soils (including agricultural soils, grasslands, forests, wetlands, and permafrost) have revealed a high diversity of DNA and RNA viruses. The majority are bacteriophages, along with viruses of archaea and eukaryotes. One of the central questions in soil viral ecology is how viruses influence key metabolic pathways that are prevalent in soil. Intriguingly, many soil viruses encode putative AMGs that could affect carbon, nitrogen, phosphorus, or sulfur nutrient transformations in soils[49,51–56]. For example, soil viromes contain genes for carbohydrate-active enzymes such as cellulases, chitinases, and lignin-degrading enzymes, which could influence the breakdown of organic matter in soils where they are enriched, such as in wetland and peatland soils[39,46,52–54]. Some soil viruses have also been found to encode genes involved in endospore formation[52], suggesting that viral dark matter in soil may be involved in host stress responses as well as metabolism. However, these functional annotations are based on sequence homology and remain mostly unvalidated. Researchers have predicted these viral genes might augment microbial metabolism in soil, but direct biochemical evidence of their activity is largely lacking. Nevertheless, their presence suggests that soil viruses might enhance their hosts' abilities to utilize complex nutrients or survive harsh soil conditions, thereby contributing to processes like decomposition and nutrient turnover, if these viral genes are indeed expressed and functional during infection.

In a recent compelling case confirming the activity of AMGs, Wu *et al.* (2022) isolated several candidate viral chitosanase genes from soil metagenomes and experimentally demonstrated their activity[54]. Chitosanases break down chitin derivatives common in fungal cell walls and insect exoskeletons. Multiple viral chitosanase AMGs were expressed, and one gene product showed clear endo-chitosanase activity, confirming it as a functional enzyme. The researchers then crystallized the viral enzyme and solved its structure at ultrahigh resolution, providing a rare atomic structure of a soil viral AMG product[54]. Further evidence comes from soils contaminated with organochlorine pesticides (OCPs). Zheng *et al.* (2022) demonstrated that viral genomes in OCP-contaminated soils harbored a higher abundance and diversity of AVGs associated specifically with pesticide degradation[57]. Among these was a viral-encoded L-2-haloacid dehalogenase, which was experimentally validated to degrade OCP precursors, thereby alleviating pesticide toxicity and improving host bacterial growth[57].

These works provide proof-of-concept that viral dark matter in soils can be experimentally illuminated: genes predicted from metagenomes can yield active proteins with biogeochemical roles. They also underscore how little we have explored, with each study noting that their characterized enzymes were among the very few soil virus AVGs to ever be biochemically characterized. In summary, soil studies hint at diverse viral contributions to soil biochemistry (Figure 2), but apart from rare attempts to study them in detail like the viral chitosanase and dehalogenase enzymes, most of these functions remain putative until confirmed by biochemical assays.

### 3.3. Aquatic Environments

Freshwater and marine ecosystems host enormous diversity of phytoplankton-infecting viruses, including bacteriophages as well as viruses of protists. In fact, much of what we know about AVGs comes from aquatic systems, especially the oceans. Early examples of AMGs came from marine ecosystems: as mentioned, cyanophages carry photosynthesis genes (*psbA*, *psbD*, *hli*, etc.)[4,58–60], presumably to boost their host's photosynthetic output during infection and thus increase energy availability for phage production. These photosynthetic AMGs are actively expressed, for instance, phage *psbA* transcripts increase during *Prochlorococcus* infection[58] and have been shown to enhance host recovery from photoinhibition. Similarly, marine phages infecting autotrophs carry genes for nutrient acquisition. Cyanophages, for example, encode phosphate transporter genes like *pstS* to help their hosts scavenge phosphate[61], and viruses infecting ammonia-oxidizing archaea have been found to encode ammonia monooxygenase subunits (amoC), with viral copies being highly abundant and actively expressed in metagenomic samples[62], linking viral infection to the nitrogen cycle.

Marine viruses also encode enzymes in core metabolic pathways such as glycolysis or the TCA cycle[6], highlighting viruses' broader metabolic reprogramming potential in central carbon metabolism. Recently, Tian *et al.* (2024) systematically cataloged marine viral AMGs from *Tara* Oceans[63], a planetary-scale metagenomic sequence database sampled from diverse ocean ecosystems, identifying over 86,000 AMGs grouped into nearly 23,000 gene clusters, with 32% of the clusters having no matches in existing databases at the time[14]. They mapped these AMGs to 128 metabolic pathways, highlighting lipid, nucleotide, and carbohydrate metabolism pathways as "hot spots" where viruses gene copies either outnumbered their cellular counterparts or otherwise contributed to the majority of steps required for metabolic processes[14]. Additionally, Zayed *et al.* (2021) conducted a detailed analysis of marine viruses through an expanded viral HMM profile database (*efam*) and revealed tens of thousands of novel viral protein families, most lacking known functional annotations[64]. This database, enriched by marine viral metaproteomic data, doubled the functional annotation rate of viral proteins compared to conventional methods, providing a powerful resource for future marine viral dark matter studies.

In freshwater systems, long-term studies further expand the scope of known viral AMGs. For example, Zhou *et al.* (2025) characterized over 1.3 million viral genomes across a 20-year time series in Lake Mendota (Wisconsin, USA), identifying 574 AMG families, including genes involved in photosynthesis (*psbA*), methane oxidation (*pmoC*), and hydrogen peroxide decomposition (*katG*)[15]. Many of these AMGs were consistently active, suggesting stable roles in freshwater microbial metabolism over decades, despite substantial changes in environmental conditions and community composition[15]. However, like marine ecosystems, most viral genomes and their encoded proteins in this study remained uncharacterized. Likewise, despite substantial progress in studying aquatic viruses, many marine viral genes are still uncharacterized due to the explosion of viral genes and genomes sequenced from the global oceans. For instance, Gregory *et al*. (2019) expanded marine viromes to include nearly 200,000 viral populations globally, yet most remained functionally unknown[65]. Collectively, aquatic studies highlight that while some AMGs have been experimentally characterized and linked clearly to ecological processes (Figure 2), the vast majority of viral-encoded proteins remain uncharacterized, underscoring the magnitude of viral dark matter and its global significance.

### 3.4. Extreme Environments

Viruses thrive even in extreme environments such as hot acid springs, hypersaline lakes, deep sea hydrothermal vents, often infecting extremophiles and carrying unusual genes adapted to harsh conditions[66]. These extreme microbiomes are rich in viral dark matter, partly because the host organisms themselves are genetically distant from well-studied model species[66–68]. Viruses from these habitats encode proteins with distinct biochemical adaptations, such as thermally stable DNA polymerases and unique structural proteins capable of functioning under extreme temperatures, acidity, and salt concentrations[68,69].

Thermal environments, particularly hot springs such as those in Yellowstone National Park (USA), provide striking examples of viruses uniquely adapted to extreme conditions. Metagenomic and single-cell genomic analyses have revealed extensive and complex networks of virus-host interactions in Yellowstone's hot springs, where over 60% of microbial cells harbored viruses, often hosting multiple distinct viral types simultaneously[70]. Many of these viruses infect thermophilic archaea like *Sulfolobus islandicus* and exhibit remarkable genomic diversity[68]. Intriguingly, these viruses frequently encode toxin-antitoxin systems, whereby viral infection confers competitive advantages to their hosts by killing competing microbes[12]. For instance, chronic infections by *Sulfolobus* spindle-shaped viruses (SSVs) mediate host fitness through virus-encoded toxins that specifically target uninfected, CRISPR-immune populations of competing archaea, illustrating a form of virus-host mutualism that enables coexistence and shapes microbial community structure[12,71].

Hypersaline and hyperarid environments, such as the Atacama Desert (Chile) and the Great Salt Lake (Utah, USA), also harbor viruses that significantly influence microbiomes through encoded metabolic and stress-response genes. In microbial consortia inhabiting halite nodules of the hyper-arid Atacama Desert, viruses infect diverse archaea and bacteria. Crits-Christoph *et al*. (2016) identified these viruses as crucial mediators of ecological interactions and microbial adaptation to high osmotic pressure[69]. Complementing these findings, Hwang *et al*. demonstrated that Atacama viruses encode genes involved in microbial stress responses and spore formation, enhancing host resilience against extreme desiccation and radiation[72]. Similarly, in hypersaline Great Salt Lake sediments, viruses carry AMGs associated with core metabolic processes, including photosynthesis, carbon fixation, formaldehyde assimilation, and nitric oxide reduction, directly linking viral genes to essential elemental and nutrient transformations[73].

Viral contributions extend even to deep-sea hydrothermal vents, another extreme habitat characterized by high temperatures, pressures, and chemical gradients. In these sulfur-rich habitats, sulfur-transforming AMGs are especially diverse and widespread. For example, Anantharaman *et al.* (2014) identified viruses infecting marine chemolithoautotrophic bacteria that encode reverse dissimilatory sulfite reductase (*rdsr*) genes, enabling the conversion of elemental sulfur into sulfite at a key bottleneck in energy metabolism through sulfur oxidation[74]. Similarly, Anderson *et al.* (2014) reported viral genomes from vent ecosystems that carry AMGs involved in sulfur and methane metabolism[75]. Kieft *et al*. (2021) further showed that AMGs such as *dsrA*, *dsrC*, *soxYZ*, and *soxCD* were not only widespread across hydrothermal ecosystems, but also highly expressed, at levels several orders of magnitude higher than in background deep-sea samples[76]. More recently, Langwig *et al*. (2025) identified thousands of uncharacterized viruses from globally distributed hydrothermal vents, many encoding AMGs involved in sulfur and nitrogen transformations[77]. These findings demonstrate that viruses directly contribute to microbial energy metabolism in hydrothermal vent ecosystems, highlighting their functional importance in extreme, sulfur-rich environments in the deep sea.

Together, these examples from thermal springs, deserts, hypersaline lakes, and deep-sea vents illustrate how extreme environments expand our understanding of viral diversity, function, and adaptation. Such ecosystems highlight the tremendous potential of viral dark matter—encoding novel proteins and biochemical capabilities adapted to extreme conditions (Figure 2). Yet, the biochemical functions of most of these viral proteins remain speculative, underscoring a pressing need for targeted experimental validation. Characterizing extremophile viral proteins could yield enzymes of substantial biotechnological interest, such as those exhibiting remarkable thermal stability, salt tolerance, or acid resistance. To harness this potential, interdisciplinary efforts combining metagenomics, structural biology, and biochemical experimentation are essential to illuminate the functional roles of viruses thriving at life's extremes.

## 4. Challenges in the Characterization of Viral Proteins

### 4.1. Lack of Homologous Sequences

Viral proteins are incredibly diverse. Many viral genes are so divergent that standard sequence similarity searches (e.g. BLAST[78]) find no meaningful hits in databases. Detecting distant evolutionary relationships from sequence alone is difficult[79,80]. Thus, a newly discovered viral protein often starts as an ORF with no known relatives, yielding no clues to function. For example, a novel phage gene might not match any domains or motifs with functional annotations in standard reference protein databases like Pfam[81], PHROG[24], VOGDB[23], eggNOG[82], or KEGG KOfam[83]. This is compounded by the fact that viruses often evolve via rapid mutation, gene shuffling, or recruiting genes from hosts and then diverging them[84,85]. Remote homology detection methods (profile HMMs, etc.) can sometimes classify these proteins into broad families, but even advanced clustering using remote homology as employed by the phage-specific PHROG database could only assign putative functions to ~50% of viral protein families[24], leaving the rest labeled as "unknown function."

### 4.2. Limited Representation in Structural Databases

Structure can sometimes reveal function, for instance, a protein might have the fold of a protease or a kinase even if its sequence did not show it. But historically, very few viral hypothetical proteins have had solved 3D structures. Most solved virus protein structures are of well-known virion components (capsids, tail fibers) or enzymes (polymerases, lysozymes) from model phages or viral pathogens. Consequently, the extensive catalog of small, functionally enigmatic phage proteins, such as anti-host factors and metabolic enzymes, remains severely underrepresented in structural databases like the Protein Data Bank (PDB)[86]. This structural gap significantly hampers functional inference via fold comparisons for viral dark matter proteins. Recently, computational approaches like AlphaFold2[87] have begun addressing this gap by predicting structures for thousands of viral proteins, as exemplified by the newly developed VFOLD database (within VOGDB)[23] and the Big Fantastic Virus Database (BFVD), both of which specifically contain viral proteins overlooked by general databases[88]. The BFVD, for instance, contains over 350,000 predicted viral protein structures, approximately 62% of which show no or very low structural similarity to existing structural databases like AlphaFold DB[89] and the PDB[86]. Moreover, until recently, searching a query against millions of predicted structures was computationally infeasible; tools like Foldseek now allow fast structure-based searches[80], but the accuracy of function inference from predicted structure still needs expert verification.

### 4.3. Misleading Homology-Based Functional Predictions

A growing concern in viral genomics is the mis-annotation of viral genes with attractive but incorrect functions. High-throughput prediction pipelines can assign enticing labels to viral ORFs that turn out to be wrong upon closer scrutiny[5]. Martin *et al.* (2025) warned that the rush to catalog AVGs (and AMGs in particular) has led to an "epidemic of mis-annotation", where functions are predicted without sufficient manual curation or evidence[5]. One prominent example is the misclassification of glycoside hydrolases (GHs) in viral genomes. Viral GH-like domains, while often annotated as polysaccharide-degrading metabolic enzymes potentially aiding host nutrition, frequently have structural or virulence roles unrelated to metabolism. For instance, phage tail fibers and baseplate proteins often incorporate GH domains to degrade host surface polysaccharides, facilitating viral entry[90,91]. Similarly, phage endolysins, enzymes responsible for host cell wall degradation at the conclusion of the phage replication cycle, can be homologous to host metabolic enzymes such as chitinases or muramidases, but their role is strictly in host cell lysis rather than in nutrient breakdown[90–92]. Structural studies have illustrated that chitinases, chitosanases, and phage lysozymes share conserved core folds despite minimal sequence similarity, highlighting the difficulty in accurately predicting their biological roles solely by homology[90,91]. Overall, this is not to say that viruses do not truly encode *bona fide* GHs for organic matter decomposition, but homology-based functional predictions alone are insufficient to discern between metabolic and essential functions for viral proteins with GH domains. Without rigorous experimental validation, such structural and virulence-associated proteins can thus be erroneously classified as metabolic genes, perpetuating misconceptions in viral ecology studies[5].

### 4.4. Dependency on Host Context

Many viral proteins act by interacting with host proteins or metabolites. Consider AVGs for example: a phage protein might redirect a host regulatory pathway by binding to a host enzyme or altering its regulation. Determining such a function might require knowledge of the host target, which in turn might not be known or easy to test without the host system. If the host is not *yet* cultivated (which is typically normal for environmental microbes), we cannot do traditional genetic experiments like knockouts or complementation to learn the protein's role. This makes it challenging to design experiments, as one might not even know what substrate or condition to test. For instance, a viral protein might only function at a specific infection stage[93], in concert with specific other viral/host factors[12], or in certain environmental conditions[30], which is hard to recapitulate in isolation. The substantial dependence on host context means that even if we can purify a viral protein, we might miss its true function if the assays don't mimic the right conditions.

### 4.5. Scale of the Problem

The vast number of sequenced and uncharacterized viral genes, now in the hundreds of millions, far exceeds our current ability to functionally characterize them. Historically, functional analyses have proceeded slowly, characterizing individual genes or phages one-by-one. Recently, high-throughput approaches such as deep mutational scanning, pooled selection assays, and CRISPR-based genome editing have begun addressing this bottleneck by systematically probing tens of thousands of phage variants simultaneously[94]. However, these methods have primarily been limited to model phages and easily cultivable hosts such as *Escherichia coli*[94]. Extending such powerful approaches to diverse environmental phages and non-model hosts remains challenging due to technical hurdles in phage-host compatibility and library construction. Nonetheless, scaling these high-throughput functional genomics methods more broadly holds promise for systematically illuminating viral dark matter on a global scale.

### 4.6. Interdisciplinary Gap

There is also a cultural and communication challenge. Microbial ecologists and bioinformaticians who discover novel viral genes may not have the expertise or resources to characterize protein function, while biochemists and molecular biologists might be unfamiliar with the significance of viral genes discovered in metagenomes. Bridging this gap requires collaboration and data sharing. Often, the labs finding viral dark matter genes are different from those equipped to do protein biochemistry. Fostering connections between these groups will be crucial to tackle viral dark matter at scale.

## 5. Strategies to Characterize Viral Proteins

Addressing viral dark matter requires a multipronged approach, combining in silico predictions with *in vitro* and *in vivo* experiments. Here, we outline strategies and workflows to bridge the gap from sequence to function (Table 2), and to encourage collaborations that can accelerate the pace of discovery.

**Table 2. Key Challenges in Characterizing Viral Proteins and Their Potential Mitigation Strategies.**

| Challenge | Summary | Mitigation Strategies |
|---|---|---|
| **Lack of Homologous Sequences** | Many viral proteins are so divergent that they have no detectable relatives in sequence databases, leaving no clues to their function. | - **Remote homology and clustering:** Group unknown proteins into families to reveal distant relationships or conserved motifs.<br><br>- **Structure-based inference:** Leverage structure-guided annotation tools that can infer function from structural features even when sequences differ. |
| **Limited Structural Data** | Very few novel viral proteins have solved 3D structures (most known structures are of common capsid or enzyme proteins), which hampers function prediction by fold comparison. | - **AI-driven structure prediction:** Use AlphaFold2 to model structures for uncharacterized viral proteins, expanding structural coverage. |
| **Misleading Homology-Based Annotations** | Automated annotations can be incorrect – viral genes may be given enticing, but wrong functions based on weak similarity. Such misannotations propagate without manual curation or evidence. | - **Stringent curation:** Treat homology-based predictions with caution. Cross-check annotations against gene context and known biology; avoid assigning definitive functions without supporting evidence.<br><br>- **Experimental validation:** Prioritize lab experiments/tests for high-impact predictions. Validating a few cases helps correct errors and improve annotation accuracy. |
| **Dependency on Host Context** | Several viral proteins may act only in the presence of their host or under specific conditions. If the host is uncultivated or the interaction is complex, it's difficult to deduce or test the protein's function in isolation. | - **Heterologous expression:** Clone and express the viral gene in a model organism or cell-free system to test its activity *in vitro*, bypassing the need for the native host.<br><br>- **Genetic complementation:** Introduce the viral gene into a host strain lacking the equivalent gene to see if it can restore the missing function, indicating a similar role. Conversely, delete the viral gene to observe effects on infection, linking the protein to a phenotype.<br><br>- **Environmental context assays:** Leverage metatranscriptomics/metaproteomics to check if the viral gene is expressed under relevant conditions in natural samples. Use microcosm |
| | | or mesocosm experiments (with labeled substrates, stable isotope probing) to see if the presence of the viral gene correlates with specific metabolic activities in a community setting. |
| **Scale of Viral Dark Matter** | The number of unknown viral genes far outpaces our capacity for one-by-one characterization. Traditional experiments are slow and usually limited to model viruses and hosts, making it infeasible to tackle the vast "viral dark matter" by conventional means alone. | - **Prioritization of targets:** Use computational analyses to triage which unknown viral proteins to study first.<br><br>- **High-throughput screening:** Develop scalable assays to probe many genes in parallel. |
| **Interdisciplinary Gaps** | A cultural/communication divide exists between those who discover viral genes (e.g., virologists, microbial ecologists, bioinformaticians) and those who characterize protein function (biochemists, molecular biologists). As a result, novel viral proteins found *in silico* often remain unstudied in the lab, simply because the expertise and data needed to investigate them reside in different communities. | - **Cross-field collaborations:** Establish dedicated programs or consortia to bridge viromics and experimental biology. By connecting virus discovery labs with protein function labs, researchers can jointly tackle top-priority viral proteins.<br><br>- **Shared resources and incentives:** Create user-friendly databases listing "most-wanted" viral proteins of unknown function, including their prevalence, predicted features, and structural models to attract interest from broader scientific communities. |

## 5.1. Prioritize and Group Unknown Proteins Using Bioinformatics

Given the enormous scale of unknown viral proteins, prioritization is essential. Bioinformatic analyses can identify candidate genes of interest based on several criteria: (a) *Abundance or ubiquity* – proteins that are highly abundant in metagenomes or occur in many samples, indicating ecological importance; (b) *Genomic context* – for example, a viral gene adjacent to known metabolic genes may also have a metabolic function; (c) *Phylogenetic or taxonomic scope* – proteins unique to viruses infecting certain hosts or environments, pointing to specialized functions; and (d) *Conservation* – unknown proteins that are conserved across many related viruses suggest an important role worth investigating.

Clustering and remote homology tools are very useful at this stage. Databases like PHROG and VOGDB group viral proteins into families and provide consensus annotations or tags (if available)[23,24]. For instance, if a particular PHROG family is enriched in unknown proteins from soil viruses and that family is widespread, that makes it a good target to study via one representative member. Likewise, knowing a protein falls in a VOGDB category of "viral proteins beneficial for host (Xh)" versus "viral replication (Xr)" vs "unknown (Xu)"[23] can guide the types of assays to consider. For example, an unknown protein classified as likely structural (based on genomic context or weak homology) would prompt tests for virion incorporation, whereas one in a host-benefit cluster might be tested for enzymatic activity affecting host metabolism or resistance.

Structural prediction and comparative approaches further enhance functional inference capabilities. Recent computational advances like AlphaFold2[87] predict high-quality three-dimensional models for many small proteins. These structural models can be searched against known protein structures using tools such as Foldseek, enabling highly sensitive detection of remote structural similarity compared to traditional sequence-based searches[80]. Specialized databases, such as BFVD[88] and VOGDB's structural database VFOLD[23] provide more structural resources. Thus, structural searches within BFVD and VFOLD could greatly increase the chances of finding structural homologs for previously unannotated viral proteins, offering substantial clues for their functional characterization.

Moreover, newly developed bioinformatics tools such as PHOLD (Phage Annotation using Protein Structures) are specifically designed to leverage these structural resources. PHOLD uses protein structural homology rather than sequence similarity alone to sensitively annotate bacteriophage genomes from metagenomes, significantly enhancing functional annotation accuracy for uncharacterized viral proteins (github.com/gbouras13/phold). Along similar lines, the recently developed Protein Set Transformer (PST) framework offers a complementary approach by modeling viral genomes as sets of proteins and learning contextualized embeddings that reflect both functional and structural relationships[95]. PST can cluster unannotated proteins based on shared structural features, even in the absence of sequence similarity, and has demonstrated strong performance in annotating viral proteins of unknown function and identifying broad functional modules from genome context[95]. Together, tools like PHOLD and PST can substantially improve prioritization of viral proteins for experimental validation by narrowing down high-confidence, structure-aware candidates.

In summary, computational tools help reduce the search space and provide testable hypotheses. When considering a novel protein for a biochemical characterization, one should ideally have some prediction for its possible function (e.g., "this phage protein has a nuclease-like fold and is conserved in many gut phages, it might be a DNase that helps the phage overcome host DNA defense"). Such a step is crucial to enable focused and tractable experimental efforts.

### 5.2. Experimental Approaches to Characterize Viral Dark Matter

Closing the gap between predicted function and confirmed activity for viral proteins will require a concerted increase in experimental work. So far, logistical challenges have limited these efforts, many viruses with interesting genes infect hosts that are difficult or improbable to culture in the lab. Nevertheless, creative approaches can be employed to experimentally probe viral protein functions without needing the full virus-host system.

For example, if a viral genome harbors a candidate enzyme gene, researchers can synthesize the gene and express it in a model organism or cell-free system to test its activity *in vitro*. Advances in heterologous expression and protein engineering now make it feasible to produce many viral enzymes, even those from rare environmental phages[96]. Successful expression opens the door to biochemical assays: does the protein catalyze the expected reaction? Is its activity measurable? One can also pursue structural biology (X-ray crystallography or cryo-EM) to glean mechanistic insights once the protein is purified. In the case of the soil viral chitosanase mentioned above, the investigators bypassed the need to culture the virus by cloning the gene, expressing the protein in *E. coli*, and determining its 3D crystal structure which confirmed the anticipated active sites for chitosan hydrolysis[54]. This approach can be broadly applied to other viral enzymes of interest.

In addition to *in vitro* biochemistry, genetic experiments can illuminate function. Viral genes suspected to influence host metabolism could be tested by introducing them into a host strain lacking the corresponding native gene (a complementation test) to see if the viral gene can restore the function[97]. For instance, if a phage carries a *folA* (dihydrofolate reductase) gene, one could knock out *folA* in the host and see if the phage-encoded version rescues growth. Similarly, one could knock out or inhibit the host's gene during viral infection to see if the virus's version compensates for the loss, demonstrating the viral protein's functionality. Where possible, constructing mutant viruses that delete the gene and observing the effect on replication and host physiology is the most direct evidence of function, though this remains technically difficult for many environmental viruses.

Modern 'omics and cell biology techniques also offer indirect routes to validation. Metatranscriptomics and metaproteomics can detect whether putative AMGs are actually expressed during infection in natural samples. For example, if a viral gene is highly expressed at the precise time it would be needed (say, a viral nitrogen metabolism gene expressed during host nitrogen starvation), that bolsters the case that it is functional in that context. Mesocosm and microcosm experiments, which maintain semi-natural systems under controlled conditions, offer a useful intermediate between lab and field studies. When paired with genomics, transcriptomics, proteomics, or stable isotope probing, they allow researchers to track viral gene expression and function following controlled environmental shifts, without needing to isolate viruses or hosts. Stable isotope probing (SIP) could be particularly valuable for testing AMG activity: by incubating a community with a labeled substrate (e.g., $^{13}CO_2$) and tracing the incorporation of the label into biomolecules in the mesocosm/microcosm, researchers can assess whether a virus population carrying a specific AMG contributes to the associated metabolic process. For instance, if a virus encodes a methanogenesis-related gene, SIP could confirm its functional role by showing label incorporation into methane or host biomass when the viral gene is present and expressed in the active community. Although challenging, similar ecosystem-level assays have recently been used to infer viral activity *in situ*[98,99].

To overcome the limitations of one-at-a-time gene characterization when faced with many promising candidates, emerging high-throughput strategies are beginning to make functional viromics more scalable and applicable to diverse environmental phages. For example, Chen *et al*. developed the PhageMaP method which enables genome-wide interrogation of phage gene function by combining modular genome engineering with pooled phenotypic screens[100], allowing researchers to map essential, nonessential, and host-specific genes across multiple phages and bacterial hosts[100]. This approach provides a flexible framework for functional analysis beyond classic model systems.

Similarly, Huss *et al*. developed the method Meta-SIFT to address the challenge of annotating proteins without known homologs by using deep mutational scanning data to identify conserved sequence motifs and functionally important residues in viral proteins[101]. By integrating this with metagenomic data, Meta-SIFT can predict meaningful functional regions even in highly divergent proteins[101]. Together, these methods offer promising paths to extend functional genomics to viral dark matter, enabling more systematic, large-scale annotation of unknown viral proteins across ecosystems.

Overall, there is a rich toolbox available, from classical biochemistry to cutting-edge multi-omics to probe viral protein function. We deliberately refrain from prescribing an exact experimental workflow, as the optimal approach will differ case by case. The key point is that integrating experimental validation into viromics studies is essential. Even a few targeted validations can have outsized impact: they provide "proof of concept" that certain viral genes are truly functional, help calibrate bioinformatic predictions, and sometimes reveal surprises that revise our understanding of viral capabilities. Going forward, collaborations between the bioinformaticians and microbial ecologists who identify candidate genes and the experimental biochemists who can test them will be especially powerful. By combining strengths, such cross-disciplinary teams can systematically chip away at the mountain of viral dark matter, one function at a time.

### 5.3. Fostering Collaboration and Data Sharing

To engage a technically diverse set of researchers for viral protein characterization, a cultural shift and new frameworks may be needed. One suggestion is to establish dedicated programs or consortium efforts for virus protein functional genomics. These could be analogous to past structural genomics initiatives which solved structures for many hypothetical proteins. For example, a project could take the top 100 most abundant unknown viral proteins (across various environments) and systematically attempt to express, assay, and solve structures for them. Results (including negative ones) should be shared in databases so others can learn which proteins were recalcitrant or which conditions yielded activity.

Another strategy is creating user-friendly databases that merge viral genomics data with information tailored for experimentalists. The IMG/VR system already links viral genomes with metadata and predicted functions[25]. Expanding such resources to include "most wanted" lists of unknown viral proteins, with their environmental distribution and any predictions or structural models, could attract researchers from outside virology. For instance, a biochemist interested in novel glycosidases could check a curated list of uncharacterized viral carbohydrate-active enzymes, complete with model predictions and alignments. The VOGDB and PHROG catalogs are a starting point, providing browsable families of viral proteins; experimentalists can use these to find targets that are not just one-off ORFans (singletons) but part of larger conserved families (so that any findings will be broadly relevant). Better integration of bioinformatic outputs with experimental databases would lower the barrier for scientists to pick up viral proteins for study.

## 6. Outlook

## 6.1. Viral Dark Matter for Basic Science

Why should experimentalists invest their time in characterizing proteins encoded by uncultivated viruses? First, from the standpoint of fundamental science, viral dark matter holds deep insights into evolution and ecology. Viruses have had billions of years to sample sequence space and evolve novel solutions to biological problems. Many protein families likely originated in viruses or have been extensively reshaped by viral evolution. By exploring viral proteins of unknown function, we may discover entirely new protein folds or biochemical activities that expand the known repertoire of life's catalysts. Such discoveries can, in turn, illuminate how viruses manipulate hosts and influence ecosystem processes. Filling in these gaps is key to accurately predicting ecological responses to change, such as in the context of human, animal, and plant diseases, climate-driven shifts in the oceans and lakes, or nutrient alterations in soil.

## 6.2. Viral Dark Matter for Biotechnology

Second, there is a strong practical incentive: viral proteins have already proven their value in biotechnology and medicine, and many more applications undoubtedly await. A historical case in point is the discovery of reverse transcriptase in retroviruses in 1970[102,103], which was a watershed moment in molecular biology. This viral enzyme, initially a puzzling novelty, became the cornerstone of recombinant DNA technology by enabling cDNA cloning and RT-PCR. Likewise, phages have yielded a treasure trove of enzymes now routinely used in the lab: DNA polymerases for PCR and DNA sequencing, DNA ligases for cloning, and RNA polymerases like T7 RNAP for *in vitro* transcription, all originate from phages[32]. In medicine, phage enzymes are being developed as novel antimicrobials. Phage endolysins are enzymatic antimicrobials that can swiftly lyse specific bacteria, including drug-resistant strains, without harming beneficial microbes[33,34]. Similarly, phage depolymerases (which are often glycosidases) show promise in breaking down bacterial biofilms in chronic infections[104]. These successes underscore that viruses encode unique biochemistry that biotechnology can leverage. In short, today's "hypothetical" viral proteins have the potential to become tomorrow's biotech workhorses or drug leads.

## 6.3. Auxiliary Viral Genes for Bioengineering Phage Therapies

One especially promising avenue emerging from viral dark matter research is the exploitation of auxiliary viral genes (AVGs) in phage therapy. Because AVGs by definition are not required for the phage to reproduce under ideal laboratory conditions, they can often be added, removed, or modified without completely destroying phage viability. This makes them attractive targets for phage bioengineering, one can potentially equip therapeutic phages with beneficial cargo genes or remove detrimental ones to create a more effective antibacterial agent.

One example is the use of phage-encoded anti-CRISPR (Acr) proteins. Anti-CRISPRs are small proteins produced by some phages (often carried in prophage elements) that inhibit the host's CRISPR-Cas immune system, thereby protecting the phage from CRISPR-based defense[105,106]. They were originally viral dark matter genes, tiny ORFs with no known function until discovered through clever screens. Dozens of Acr families have since been identified, each targeting different types of CRISPR systems[107,108]. In the wild, not all phages have *acr* genes, but those that do have a clear advantage against CRISPR-proficient bacteria. Recognizing this, researchers have begun engineering phages to carry anti-CRISPR genes to make them more effective in therapeutic contexts. For instance, Qin *et al*. (2022) engineered a lytic phage of *Pseudomonas aeruginosa* by inserting genes for AcrIF1–3 (which block the Type I-F CRISPR system of *P. aeruginosa*)[109]. The modified phage showed enhanced ability to replicate on bacteria with active CRISPR defenses and could thereby infect and kill strains that normally would fend off phages[109]. In addition, these Acr-armed phages suppressed the emergence of phage-resistant bacterial mutants in experiments and even reduced antibiotic resistance in the bacterial population[109]. This demonstrates a powerful principle: by leveraging an AVG (in this case, an anti-immunity gene), we can improve phage therapy outcomes against bacteria that were previously difficult to eradicate.

Crucially, using AVGs in engineering is often safer or more feasible than tinkering with essential phage genes. If one tried to modify a phage's capsid or replication proteins, the phage will likely lose viability. In contrast, adding a new AVG or swapping one AVG for another can often yield a viable phage, because these genes are not strictly required for the phage life cycle in lab conditions. For example, phages can typically tolerate the insertion of a small gene like an anti-CRISPR or an anti-toxin gene in their genome, as long as careful design ensures it doesn't disrupt other elements. Deleting an AVG that encodes a detrimental function (like a phage-encoded toxin that is undesirable in a therapy phage) can also be done without rendering the phage inert, since the gene is not needed for basic replication.

As phage therapy experiences a modern renaissance due to rising antibiotic resistance, phage engineering with AVGs stands out as a way to create "smarter" therapeutics. Phage "cocktails" are commonly used to deliver multiple phage strains over the course of treatment[110,111]. We could envision, for instance, a phage cocktail for a *P. aeruginosa* infection where each phage is armed with a different AVG: one carries an anti-CRISPR to overcome bacterial immunity[109], another carries a depolymerase to dissolve biofilm matrices[104], and a third carries a miniature CRISPR array to eliminate an antibiotic resistance plasmid[112]. Such multi-pronged attacks, delivered by viruses that naturally target the bacteria, could be extremely potent. Of course, regulatory and safety considerations are paramount, any engineered phage used clinically must be well-characterized to ensure it doesn't carry harmful genes (like antibiotic resistance genes or potent toxins), compounding the need to characterize phage proteins. In this regard, focusing on auxiliary genes is ideal, because one can theoretically remove any that are problematic and insert only those that benefit the therapeutic outcome.

In summary, auxiliary viral genes represent a treasure trove of functions that viruses have evolved to give themselves an edge, and we can repurpose those functions for human needs (Figure 3). Continued discovery and characterization of AVGs, including putative anti-defense proteins, will expand the catalog of techniques available for engineering phages as precision antimicrobials. This is a prime example of how decoding viral dark matter can translate into tangible applications: turning obscure phage genes into tools to fight disease.

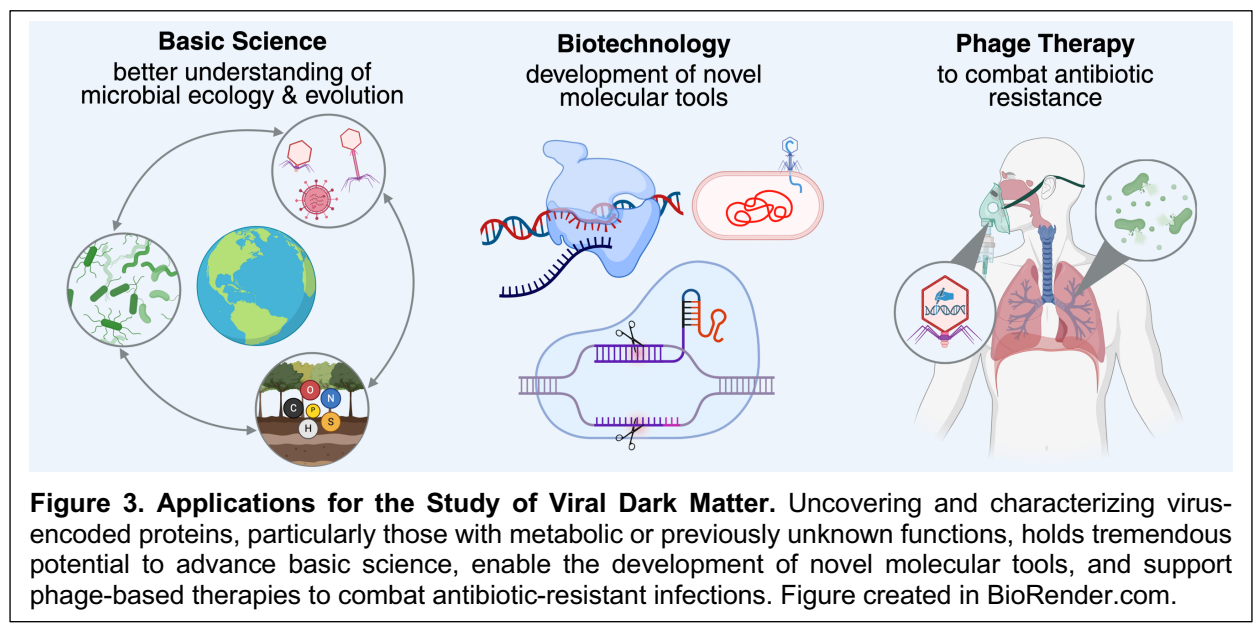

**Figure 3. Applications for the Study of Viral Dark Matter.** Uncovering and characterizing virus-encoded proteins, particularly those with metabolic or previously unknown functions, holds tremendous potential to advance basic science, enable the development of novel molecular tools, and support phage-based therapies to combat antibiotic-resistant infections. Figure created in BioRender.com.

## 7. Call to Action: Harnessing Viral Dark Matter

Given these possibilities detailed above, we urge the scientific community, especially those with biochemical and molecular biology expertise, to engage with the challenge of viral dark matter. This is a call to broaden our perspective: viruses are not just vectors of disease or abstract sequences in databases, but also reservoirs of unexplored functions. By partnering with virologists, microbial ecologists, and bioinformaticians, experimental scientists can help turn putative annotations into proven activities. The payoff includes not only advancing our basic understanding of viral ecology but also unearthing novel enzymes, therapeutics, and tools. Ultimately, shining a light on viral dark matter will enrich biology as a whole, revealing new facets of how life operates at the molecular level and providing fresh inspiration for innovation in biotechnology and medicine. The time is ripe to bring the hidden world of viral proteins into clear view, and the benefits of doing so will resonate across scientific disciplines. Let us answer this call to action and illuminate the unknown viral functions that have waited too long in the shadows.

## 8. Acknowledgements

We thank members of the Anantharaman lab for helpful discussions. JCK was funded by a National Science Foundation Graduate Research Fellowship. This work was supported by National Institute of General Medical Sciences of the National Institutes of Health under award number R35GM143024 (to KA).